\documentclass{article}
\usepackage{graphicx}
\usepackage{amsmath}

\input{tcilatex}

\begin{document}

\title{Quantum Field Theory of the Pinned Density Wave}
\author{J.H. Miller, Jr.,$^{a,b}$ C. Ord\'{o}\~{n}ez,$^{a,c}$ and E. Prodan$^{a}$ \\
a) Department of Physics,\\
b) Texas Center for Superconductivity\\
c) World Laboratory Center for Pan-\\
American Collaboration in Science and\\
Technology\\
University of Houston, Houston, Texas, 77204-5506}
\date{September 14, 1998}
\maketitle

\begin{abstract}
A model is discussed in which an electric field induces quantum nucleation
of soliton-antisoliton pairs in a pinned charge or spin density wave.
Coulomb blockade prevents pair creation until the electric field exceeds a
sharp threshold value, which can be much smaller than the classical
depinning field. We calculate the vacuum state energy and expectation value
of the phase $\phi $, which is treated as a quantum scalar field. We find
that the phase can also be much smaller, below threshold, than predicted by
classical ``sliding'' density wave models.
\end{abstract}


\newpage

Charge and spin density waves are examples of spontaneous symmetry breaking,
in which pairs of electrons and holes condense into a new ground state.$^{1}$
A charge density wave (CDW) forms in a linear chain compound when the
electronic charge density becomes modulated: $\rho \left( x\right) =\rho
_{0}\left( x\right) +\rho _{1}\cos \left[ 2k_{F}x-\phi \left( x,t\right) %
\right] $, where $\phi $ is the phase. A spin density wave (SDW) has a
modulated spin density, $\Delta \mathbf{S}\left( x\right) =\Delta \mathbf{S}%
_{0}\cos \left[ 2k_{F}x-\phi \left( x,t\right) \right] $, and is equivalent
to two out-of-phase CDWs for the spin-up and spin-down subbands. Although
pinned by impurities, a density wave (DW) can transport a current when an
applied field exceeds a threshold value $E_{T}$. The simplest model of a
pinned DW is a sine-Gordon (s-G) model,$^{2}$ in which the ''vacuum states''
are the $2\pi n$ minima about which a DW would oscillate at the pinning
frequency $\omega _{0}$ in the absence of dissipation. Finite wavelength
modes, or phasons, can propagate at the phason velocity, $c_{0}=\mu
^{-1/2}v_{F}$, where $\mu =M_{F}/m_{e}$ is the Fr\"{o}hlich mass ratio.

The Hamiltonian for a pinned DW, including Coulomb interactions, can be
written as: 
\begin{equation}
H=\int dx\left\{ \dfrac{\pi ^2}{2D} +\dfrac 12 Dc_0^2\left( \dfrac{\partial
\phi }{\partial x} \right) ^2+D\omega _0^2\left[ 1-\cos \phi \right] +u_E%
\left[ \phi -\theta \right] ^2\right\} \text{,}  \tag{1}
\end{equation}
where $D=\dfrac{\mu \hbar }{4\pi v_F}$ (per spin per chain),$^3$ and the
canonical momentum density is given by $\pi =D\partial _t\phi $. The last
term in Eq. (1), introduced by Krive and Rozhavsky,$^4$ is the electrostatic
energy density $\tfrac 12\varepsilon \left( E_{int}\pm E\right) ^2$ due to
the applied field $E$ and the internal fields $E_{int}\sim -\left( \phi
/2\pi \right) E^{*}$ generated by phase variations. Here, $E^{*}$ represents
the internal field produced by a $2\pi $ soliton-antisoliton (S-S$^{\prime }$%
) pair (Fig. 1(a)), $u_E=\tfrac 12\varepsilon A\left( E^{*}/2\pi \right) ^2$
is the electrostatic energy density, $\theta =2\pi E/E^{*}$ is proportional
to the applied electric field, and $\phi $ is measured with respect to its
value at $x=$ $\pm \infty $. The above Hamiltonian is equivalent to the
bosonic form of the massive Schwinger model, as discussed by S. Coleman.$^5$
A topological soliton (antisoliton) of width $\lambda _\phi =c_0/\omega
_0\sim 1\mu m$ $\left( >>\lambda _{dw}\right) $ carries a charge $\mp
e^{*}=\mp n^{\prime }e$ per spin-chain, where $n^{\prime }$ is the
condensate fraction. A soliton-antisoliton (S-S$^{\prime }$) pair can
nucleate by quantum tunneling when an electric field is applied.$^6$ In a
related picture, a pair is created when a quantum soliton Zener tunnels
through a ``Bardeen pinning gap.''$^{7,3}$

The electrostatic (Coulomb) energy of the S-S$^{\prime }$ pair introduces a
sharply defined threshold field for pair creation. Figure 1(a) shows an S-S$%
^{\prime }$ pair, displaying the phase $\phi $ and excess charge density $%
\delta \rho =-\left( e^{*}/2\pi \right) \partial _{x}\phi $ as functions of
position. The pair, analogous to a parallel plate capacitor of separation $L$
and cross-sectional area $A$ per chain (per spin), produces an internal
field of magnitude $E^{*}=e^{*}/\varepsilon A$. The enormous dielectric
constant, $\varepsilon \sim 10^{8}\varepsilon _{0}$, may include an
intrinsic contribution $\varepsilon _{DW}$ from the pinned DW, as well as a
substantial contribution $\varepsilon _{s}$ due to screening by the normal
carriers. When an external field $E$ is applied, the difference between the
electrostatic energy of a state with a pair and that of the vacuum is given
by $\Delta U=\dfrac{1}{2}\varepsilon AL\left[ \left( E\pm E^{*}\right)
^{2}-E^{2}\right] =e^{*}L\left[ \dfrac{1}{2}E^{*}\pm E\right] $. Note that $%
\Delta U$ is positive when $\left| E\right| <\dfrac{1}{2}E^{*}$, so
conservation of energy forbids the vacuum to produce a pair for fields less
than a threshold value $E_{T}\equiv \dfrac{1}{2}E^{*}$ (corresponding to $%
\theta =\pi $). The threshold voltage across a region of length $L$ will
thus be $e^{*}/2C$ (where $C=\varepsilon A/L$),$^{3}$ by analogy to Coulomb
blockade in tunnel junctions. Above threshold, the S-S$^{\prime }$
nucleation and annihilation events will become correlated in time, by
analogy to time-correlated single electron tunneling (Fig.1(b)).$^{8}$

This model has at least two important consequences. First, the threshold
field for pair creation, $E_{T}\left( =e^{*}/2\varepsilon A\right) $, can be
much smaller than the classical depinning field in the s-G model, $%
E_{c}\equiv 2\pi D\omega _{0}^{2}/e^{*}$, provided $u_{E}<D\omega _{0}^{2}$.
Second, the phase displacement $\phi $ below threshold can also be much
smaller than either the classical s-G prediction, $\phi =\sin ^{-1}\left(
E/E_{c}\right) $, or the substantial phase displacements predicted by more
realistic classical models.$^{9,10}$ A growing body of evidence demonstrates
that the CDW phase displacements below threshold are extremely small for a
wide range of temperatures in NbSe$_{3}$ and TaS$_{3}-$a fact which cannot
readily be interpreted classically. This evidence includes: (1) the observed
2$^{\circ }$ displacement of CDW phase in NbSe$_{3}$, as determined by NMR
experiments,$^{11}$ when $E=\left( 3/4\right) E_{T}$, (2) the complete
absence of any increase in CDW dielectric response (``critical
polarization'') below threshold in NbSe$_{3}$ $^{12}$ and TaS$_{3}$,$^{13}$
in serious contradiction with classical predictions,$^{9,10}$ and (3) most
recently, the observed absence of any change in CDW wavevector $Q$ near the
contacts below threshold.$^{14}$

The primary goals of this paper are, treating the phase $\phi $ as a scalar
quantum field, to calculate the vacuum state energy, $\varepsilon $, and
expectation value of the phase $\left\langle \phi \right\rangle $ in the
metastable regime. In order to evaluate $\varepsilon $ and $\left\langle
\phi \right\rangle $, we follow Coleman$^{15}$ and employ a variational
method, choosing the trial states to be coherent states. Such a choice is
especially appropriate for a density wave, whose actual ground state (of the
3-D system) is, in fact, a coherent state. A CDW, for example, consists of a
condensate of electron-hole pairs coupled to a condensate of $2k_{F}$
phonons, which are bosons. Such a ground state is characterized by a
well-defined expectation value $\left\langle \phi \right\rangle $. It is
pointed out in Ref. $\left[ 15\right] $ that a piece of every term in the
perturbation series expansion is effectively included by this variational
approach.

Rescaling $\left( ct,x\right) =\left( x^{0},x^{1}\right) \rightarrow \dfrac{1%
}{Dc_{0}^{2}}\left( x^{0},x^{1}\right) $, the Hamiltonian Eq. (1) now reads: 
\begin{equation}
H=\int dx^{1}\,N_{m_{0}}\left\{ \dfrac{1}{2} \left[ \left( \dfrac{\partial
\phi }{\partial x^{0}} \right) ^{2}+\left( \dfrac{\partial \phi }{\partial
x^{1}} \right) ^{2}\right] +U\left( \phi \right) \right\} =\int dx\,\mathcal{%
H}\left( x\right) \text{,}  \tag{2}
\end{equation}
with 
\begin{equation}
U\left( \phi \right) =\omega ^{2}\left[ 1-\cos \phi \right] +\dfrac{1}{2}
m_{0}^{2}\left( \phi -\theta \right) ^{2}\text{,}  \tag{3}
\end{equation}
where $\omega ^{2}=D^{2}c_{0}^{2}\omega _{0}^{2}$, $m_{0}\equiv \left[
2Dc_{0}^{2}u_{E}\right] ^{1/2}$ and $N_{m_{0}}$ means normal order with
respect to $m_{0}$. The parameter $m_{0}$ can be interpreted as a ``free
field'' mass in the absence of the washboard pinning potential.

For the theory defined by Eq. $\left( 2\right) $, all ultraviolet
divergences can be removed by normal-ordering in the interaction picture.
However, variational calculations are carried out in the Schr\"{o}dinger
picture. Nevertheless, as pointed out in Ref. $\left[ 15\right] $, one can
still define a normal order $N_m$ with respect to an arbitrary mass
parameter $m$. Following Ref. $\left[ 15\right] $, the Hamiltonian density
can be re-written as: 
\begin{equation}
\mathcal{H}=N_{m_0}\left[ \mathcal{H}_0+U\left( \phi \right) \right] =N_m%
\left[ \mathcal{H}_0+U\left( \phi ,m\right) \right]  \tag{4}
\end{equation}
where $\mathcal{H}_0$ is the kinetic piece of the Hamiltonian Eq. $\left(
2\right) $ and $U\left( \phi ,m\right) $ is defined by: 
\begin{equation}
U\left( \phi ,m\right) =\exp \left[ \frac 1{8\pi }\ln \left( \frac{m_0^2}{m^2%
}\right) \frac{d^2}{d\phi ^2}\right] U\left( \phi \right) +\frac{\left(
m^2-m_0^2\right) }{8\pi }\text{.}  \tag{5}
\end{equation}
Now, it is straightforward to calculate the expectation value of the
Hamiltonian density on the vacuum states appropriate to free fields of
arbitrary mass $m$, which is given by the free coefficient of Eq. $\left(
5\right) $. Obviously, this class of states is not rich enough to be a good
trial for the variational method. In order to ensure that $\left\langle \phi
\right\rangle $ is well defined and to allow nonzero values of $\left\langle
\phi \right\rangle $, one introduces a family of coherent states, $\left|
\xi \left( x\right) ,p\left( x\right) \right\rangle $, labeled by two
arbitrary functions of space: $\xi \left( x\right) $ and $p\left( x\right) $%
. Such coherent states have the property that, for any function of $\pi
\left( x\right) $ and $\phi \left( x\right) $, $F\left( \pi \left( x\right)
,\phi \left( x\right) \right) $, the expectation value is given by: 
\begin{equation}
\left\langle \xi \left( x\right) ,p\left( x\right) \left| N_mF\left( \hat{\pi%
}\left( x\right) ,\hat{\phi}\left( x\right) \right) \right| \xi \left(
x\right) ,p\left( x\right) \right\rangle =F\left( p\left( x\right) ,\xi
\left( x\right) \right) \text{.}  \tag{6}
\end{equation}
These are used as trial states to minimize the expectation value $%
\left\langle \mathcal{H}\right\rangle $. Using Eq. $\left( 4\right) $ and
the property Eq. $\left( 6\right) $ we have then: 
\begin{equation}
\left\langle \xi \left( x\right) ,p\left( x\right) \left| \mathcal{H}\left(
x\right) \right| \xi \left( x\right) ,p\left( x\right) \right\rangle =\tfrac
12 p\left( x\right) ^2+\tfrac 12 \left( \partial _1\xi \left( x\right)
\right) ^2+U\left( \xi \left( x\right) ,m\right) \text{.}  \tag{7}
\end{equation}
Choosing $p\equiv 0$ and $\xi \left( x\right) =\xi (=\left\langle \phi
\right\rangle )=const.$ we minimize the first two terms. The energy density
becomes: 
\begin{equation}
\varepsilon =\omega ^2\left[ 1-\left( \dfrac{m^2}{m_0^2} \right) ^{\dfrac
1{8\pi } }\cos \xi \right] +\dfrac{m_0^2}2 \left( \xi -\theta \right) ^2-%
\dfrac{m_0^2}{8\pi } \ln \dfrac{m^2}{m_0^2} +\dfrac{m^2-m_0^2}{8\pi } . 
\tag{8}
\end{equation}
Minimizing $\varepsilon $ with respect to $m$ and $\xi $, one obtains the
following self-consistency equations:

\begin{equation}
\left\{ 
\begin{array}{l}
\dfrac{m^2}{m_0^2} =\allowbreak \dfrac{\omega ^2}{m_0^2} \left( \dfrac{m^2}{%
m_0^2} \right) ^{\frac 1{8\pi }}\cos \xi +1, \\ 
\\ 
\allowbreak \dfrac{\omega ^2}{m_0^2} \left( \dfrac{m^2}{m_0^2} \right)
^{\frac 1{8\pi }}\sin \xi +\xi -\theta =0.
\end{array}
\right.  \tag{9}
\end{equation}
We have also derived the above self-consistency equations by extending the
mean field approximation first developed by Glimm, Jaffe, and Spencer$^{16}$
for quartic interactions and then extended by Imbrie$^{17}$ to polynomial
interactions. Details of our calculations will be reported elsewhere.$^{18}$

Equations (9) enable one to determine both the mass-ratio $m/m_0$ and phase
displacement $\xi \equiv \left\langle \phi \right\rangle $ self-consistently
in the metastable regime. Figure 2 shows the resulting plots of $%
\left\langle \phi \right\rangle $ vs. $\theta $ (where $\theta $ is the
normalized applied electric field) for several different values of the
parameter $\tau =\dfrac{2u_E}{D\omega _0^2}$, which represents the ratio of
the electrostatic energy to the pinning energy. The solid lines through the
symbols represent the lowest energy states. Recall that S-S$^{\prime }$ pair
creation can occur only when $\theta >\pi $ (i.e. $E>E_T=E^{*}/2$). If $%
\theta =\pi $ was taken to represent the classical depinning field (i.e. if $%
E_T$ was assumed to be equal to $E_c$), then the phase below threshold would
be given by $\left\langle \phi \right\rangle =\sin ^{-1}\left[ E/E_c\right]
=\sin ^{-1}\left[ \theta _c/\pi \right] $ according to the classical
sine-Gordon model (as indicated by the x'ed line in Fig.2). However, when $%
\tau <<1$, the phase displacements can be much smaller than these classical
s-G predictions, right up to the critical value $\theta =\pi $ above which
an instanton transition can occur without violating conservation of energy.
(Note that $E_T<<E_c$ if $\tau <<1$.) The square in Fig. 2 represents the
experimental result reported in Ref. [11], which is consistent with our
calculations, provided we take the pinning energy $D\omega _0^2$ to be much
larger than the electrostatic energy $u_E$ ($\tau \approx 0.02$ in this
case). When $D\omega _0^2>>u_E$ the threshold field for pair creation $E_T$
will be much smaller than the classical pinning field $E_c$. The observed
small phase displacements indicated by Refs. [11-14] are all consistent with 
$E_T$ being small compared to $E_c$.

Figure 3 shows plots of energy density obtained using Eg. (8) vs. $\theta $
for a case where the pinning energy dominates over the electrostatic energy
(i.e. for $\tau =0.2$). From Fig. 3, one can immediately see that the
instanton transition is forbidden unless $\theta >\pi $. When the applied
electric field is below threshold ($\theta <\pi $), pair creation is
prevented by Coulomb blockade. One can gain additional insight into the
problem by considering the case when the applied field is equal to its
critical value, i.e. $\theta =\pi $. Figure 4 shows plots, obtained from
Eqs. (9), of $\left\langle \phi \right\rangle $ and the mass ratio $m/m_{0}$
as functions of the parameter $\tau $ for the case where $\theta =\pi $.
Note the bifurcation that occurs when $\tau \approx 0.87$. Thus, more than
one metastable state will exist only when $\tau $ is less than this critical
value (i.e. when the pinning energy is greater than the electrostatic
energy).

The transition temperatures for DW formation can be quite high, well above
200 K in a number of materials. The coupling between parallel DW chains and
resulting 3-D coherence must therefore suppress thermal excitations of the
S-S$^{\prime }$ pairs, whose energy would be extremely small if only a
single transverse degree of freedom was considered. Ref. [3] provides
heuristic scaling arguments on how thermal excitations might be suppressed
without suppressing the S-S$^{\prime }$ tunneling amplitude for an
intermediate range of coupling strengths, where the CDW electrons remain
delocalized in real space. The situation here may be analogous to Josephson
tunneling, where individual Cooper pairs tunnel through an insulating
barrier even though the condensed pairs comprise a condensate with a single
thermal degree of freedom. A complete theoretical treatment of DW transport
should consider many interacting scalar fields $\phi _{n}$ and, perhaps,
generalize the concept of coherent Josephson-like tunneling. The
implications of a collective quantum mechanism of DW depinning are
potentially profound and far-reaching, so further work is warranted to
experimentally test the predictions here, and to further develop a DW
depinning model based on the principles of quantum field theory.

The authors acknowledge extremely helpful conversations with J. McCarten.
This work was supported, in part, by the State of Texas through the Texas
Center for Superconductivity at the University of Houston and the Texas
Higher Educational Coordinating Board Advanced Research and Advanced
Technology Programs, by the Robert A Welch Foundation (E-1221), and by the
World Laboratory Center for Pan-American Collaboration in Science and
Technology.

\textbf{Fig.1} (a) A materialized soliton-antisoliton pair, showing the
position- dependent phase $\phi \left( x\right) $, excess charge density $%
\delta \rho \left( x\right) =-\left( e^{\ast }/2\pi \right) \partial \phi /x$%
, and internal field $E^{\ast }=e^{\ast }/\varepsilon A$. (b) The
time-evolution of the phase $\phi \left( x,t\right) $, illustrating the
nucleation and subsequent annihilation of three pairs during the first
cycle. Nucleation of additional pairs is blocked in the lightly shaded
regions (due to the internal field, -$E^{\ast }$) until all pairs have
annihilated, thus leading to time-correlated pair creation and annihilation.

\smallskip

\textbf{Fig.2 }$\left\langle \phi \right\rangle $ vs. $\theta $ for $\tau
=0.02$ (stars), 0.2 (circles), 0.87 (dashed line), and 1.5 (diamonds). The
solid lines through the symbols indicate the lowest energy states. The x'ed
line represents the classical sine-Gordon prediction (defining $\theta
=\theta _c\equiv \pi E/E_c$), while the square represents the experimental
results obtained by Ross et al. [11]

\smallskip

\textbf{Fig.3 }Energy vs. electric field ($\theta $) for $\tau =0.2$,
showing the two branches corresponding to $\left\langle \phi \right\rangle
\approx 0$ (stars) and $\left\langle \phi \right\rangle \approx 2\pi $
(circles).

\smallskip

\textbf{Fig.4 }$\left\langle \phi \right\rangle $ vs. $\tau $ (left hand
axis, circles), and mass ratio $m/m_{0}$ vs. $\tau $ (right-hand axis,
stars) for $\theta =\pi $. Note that the system bifurcates when $\tau
\approx 0.87$.

\FRAME{itbpF}{3.442in}{2.4474in}{0in}{}{}{}{\special{language "Scientific
Word";type "GRAPHIC";display "PICT";valid_file "F";width 3.442in;height
2.4474in;depth 0in;original-width 3.3961in;original-height 2.4059in;cropleft
"0";croptop "1";cropright "1";cropbottom "0";filename
'fig1a.gif';file-properties "XNPEU";}}

\FRAME{itbpF}{2.9023in}{1.8671in}{0in}{}{}{}{\special{language "Scientific
Word";type "GRAPHIC";display "PICT";valid_file "F";width 2.9023in;height
1.8671in;depth 0in;original-width 1.7288in;original-height 1.2185in;cropleft
"0";croptop "1";cropright "1";cropbottom "0";filename
'fig1b.gif';file-properties "XNPEU";}}

\FRAME{itbpF}{2.9395in}{2.3523in}{0in}{}{}{fig2.gif}{\special{language
"Scientific Word";type "GRAPHIC";display "PICT";valid_file "F";width
2.9395in;height 2.3523in;depth 0in;original-width 2.8963in;original-height
2.3125in;cropleft "0";croptop "1";cropright "1";cropbottom "0";filename
'fig2.gif';file-properties "XNPEU";}}

\FRAME{itbpF}{2.7095in}{2.3004in}{0in}{}{}{fig3.gif}{\special{language
"Scientific Word";type "GRAPHIC";display "PICT";valid_file "F";width
2.7095in;height 2.3004in;depth 0in;original-width 2.6671in;original-height
2.2606in;cropleft "0";croptop "1";cropright "1";cropbottom "0";filename
'fig3.gif';file-properties "XNPEU";}}

\FRAME{itbpF}{2.9499in}{2.3947in}{0in}{}{}{fig4.gif}{\special{language
"Scientific Word";type "GRAPHIC";display "PICT";valid_file "F";width
2.9499in;height 2.3947in;depth 0in;original-width 2.9066in;original-height
2.354in;cropleft "0";croptop "1";cropright "1";cropbottom "0";filename
'fig4.gif';file-properties "XNPEU";}}

\end{document}